\documentclass[12pt,english]{article}
\usepackage{graphicx}
\usepackage{dcolumn}
\usepackage{bm}
\usepackage{comment}
\usepackage[usenames,dvipsnames]{color}
\DeclareGraphicsExtensions{.pdf,.png,.jpg}
\usepackage{anysize}
\marginsize{2.8cm}{2.8cm}{3cm}{3cm}
\usepackage{multirow}
\usepackage{amsmath, amsfonts, amssymb, mathrsfs}
\usepackage{authblk}

\date{}                     

\title{Non proportionality dependence on shaping time}

\author[1,2]{M.~Beretta}
\author[1,2]{S.~Capelli}
\author[1,2]{L.~Gironi}
\author[2]{E.~Previtali}
\author[1,2]{M.~Sisti}  

\affil[1]{Dipartimento di Fisica - Universit\`{a} di Milano-Bicocca, Milano, Italy}
\affil[2]{INFN - Sezione di Milano-Bicocca, Milano, Italy}

\begin{document}
    
\maketitle

\begin{abstract}
In recent years the scintillation mechanism of inorganic crystals has been extensively investigated in different studies \cite{NonLinRev}. The main issues are the \emph{non-proportionality} mechanism of the light response versus energy and its connection with the shaping time used for the scintillation signals. In this study the Compton coincidence technique has been used to measure the relative \emph{non-proportionality} of three crystals: CdWO$_{4}$, BGO, and NaI(Tl). To test the \emph{non-proportionality} dependence on shaping time, the preamplified scintillator pulses have been digitized and shaped with a digital trapezoidal filter. Since no analogic shaping occurs the majority of the scintillation components are digitized, thus avoiding major information losses. The obtained results suggest the existence of a correlation between the time constant of the scintillation emission and light yield.
\end{abstract}

\section{Introduction}
The efficiency of a crystal as a scintillator is parametrized by the light yield (LY), defined as the number of photons emitted per unit of absorbed energy. A remarkable aspect outlined by different studies \cite{NonLinRev} is that the $\gamma$/$\beta$ LY of a crystal is not constant with the energy deposited inside the lattice. This characteristic, referred to as \emph{non-proportionality}, prevents the correct reconstruction of the absorbed energy, affecting the energy calibration. Moreover, the LY dependency on energy strongly affects the intrinsic component of the scintillator energy resolution \cite{NonPropEResolutionLimits}. Events having the same nominal energy can in fact result in multiple depositions differently affected by the \emph{non-proportionality}, causing a final broadening of the energy peaks. 

In recent years great importance has been given to the study of \emph{non-proportionality}, because of its correlation with the attainable energy resolution of scintillation crystals \cite{NonLinRev,NonPropEResolution,NonPropEResolutionLimits}.

In order to fully understand this feature, different crystals have been characterized, exciting the scintillation both with $\gamma$ rays and charged particles, such as $\alpha$s, $\beta$s and ions. In order to characterize the crystal independently from the detector, the LY \emph{non-proportionality} has been measured as a relative variation with respect to the LY at the 662 keV $\gamma$ line, referred to as \emph{relative light yield} ($R_{LY}$). As the excitation energy lowers, the $R_{LY}$ increases for iodine crystals (NaI and CsI for example) and lowers for oxide crystals (LSO and BGO for example) \cite{Knoll}. Moreover the $R_{LY}$ is bigger in crystals characterized by worst energy resolution \cite{NonLinRev}: as a consequence, the \emph{non-proportionality} appears to be the fundamental limit to the attainable scintillators energy resolution. 
 
A definite theoretical explanation for the \emph{non-proportionality} is not available and in recent years three phenomena have been addressed to as possible causes \cite{NonLinRev}: the energy dependence of the relaxation channels, the Landau Fluctuations of the stopping power and the secondary electrons ($\delta$-ray) scattering. These models ascribe the observed \emph{non-proportionality} to an actual change in LY, depending on the influence of intrinsic characteristics of crystals on the electrons/holes diffusion inside the lattice. These models generally predicts the behaviour of crystals, but cannot give explanation to some particular features, such as the change of $R_{LY}$ obtained with a change in the shaping time chosen for the scintillation pulse \cite{DifferentST,NonLinRev}. In particular, the choice of larger shaping times both enhances the intrinsic energy resolution and reduces the $R_{LY}$. 

As a consequence, in this paper an in depth study of the \emph{non-proportionality} dependence on shaping time is presented, in order to understand how deep this relation is. The Compton coincidence technique (CCT) has been used to measure the $R_{LY}$ of three crystals with different time and light output characteristics (see table \ref{CProp}): CdWO$_{4}$, BGO, and NaI(Tl). The point in the application of this technique is that it allows the analysis of the crystal response to electrons \cite{CCT_1,CCT_BM}. Since the scintillation mechanism is based on the electron diffusion and recombination \cite{GammaFromEl}, the CCT allows an in depth analysis of this phenomenon.

\begin{table}[htbp]
\centering
\begin{tabular}{|c|c|c|c|c|}
\hline
Crystal  & Density [gm/cm$^3$] & LY [ph/MeV] & $\tau$ [$\mu$s] (Pulse Fraction) & $\lambda_{\text{max}}$ [nm]\\
\hline
CdWO$_{4}$	&	7.9	&	18500	&	1.1(40\%), 14.5(60\%)	 &	470 \\
\hline
NaI(Tl)	&	3.6	&	38000	&	0.028(55\%), 0.23(45\%)&	415 \\
\hline
BGO	&	7.13	&	8000	&	0.032(4\%), 0.46(64\%), 0.232(32\%)	&	480 \\
\hline
\end{tabular}
\caption[]{Crystal properties table. Data were taken from \cite{Knoll,BGOData,NaIData}.}
\label{CProp}
\end{table}

In order to test the \emph{non-proportionality} dependence on shaping time, the preamplified pulses have been digitized, allowing the subsequent application of a digital filter. Since no analogic shaping occurs, the majority of the scintillation components are digitized and major information losses are avoided. Moreover, the application of different shaping times on the same digitized pulses avoids possible differences in the analysis output due to the acquisition of different measurements.

In the next sections the experimental setup will be described, alongside with the applied analysis techniques.

\section{Experimental setup}
\begin{figure}[htbp]
\centering
\includegraphics[width=.5\textwidth,origin=c]{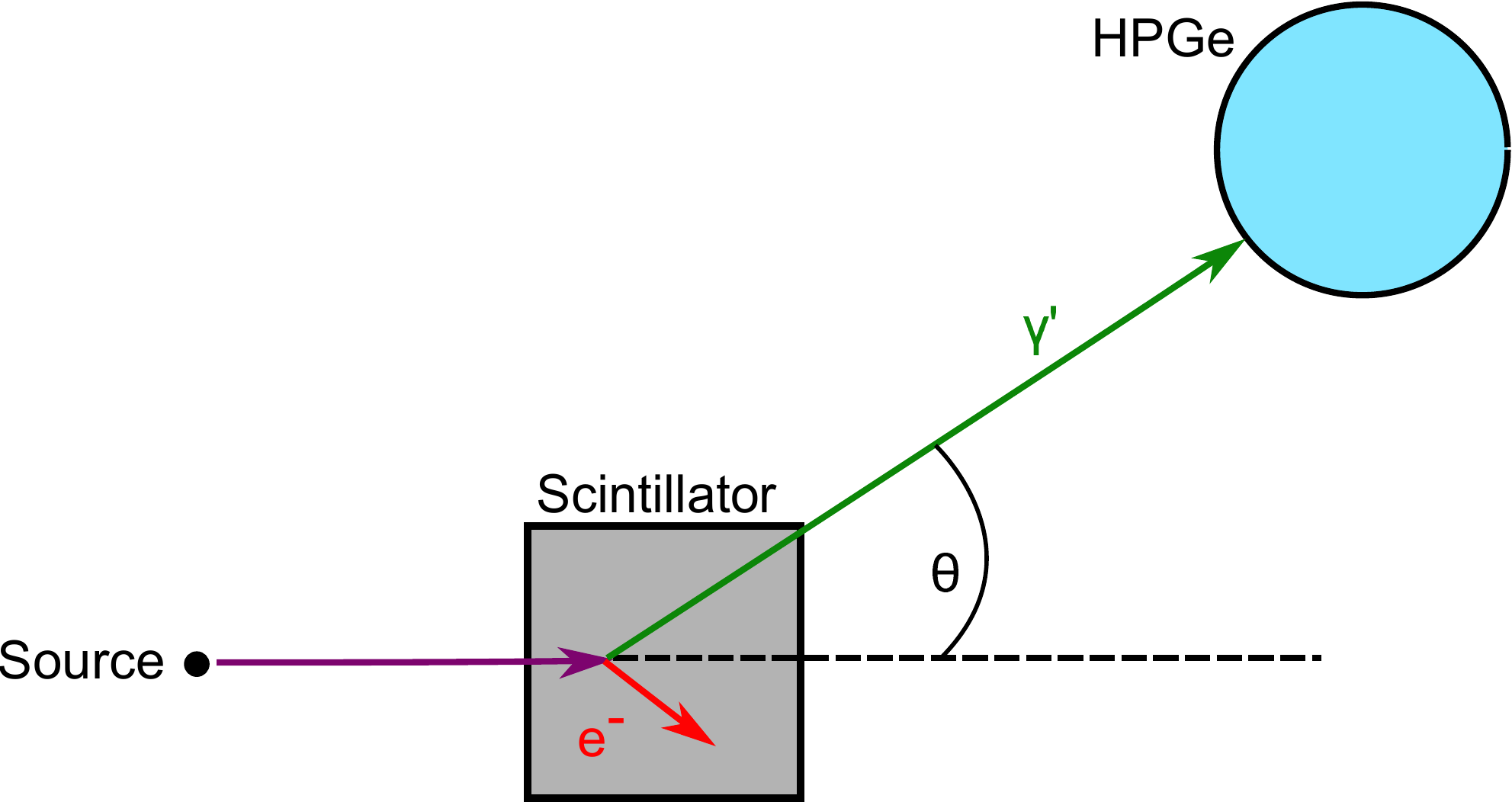}

\caption[CCT diagram]{Scheme of the CCT experimental setup. Events in scintillator are recorded only when a Compton scattered photon ($\gamma$') is simultaneously detected by HPGe, located at a fixed angle $\theta$ off of the axis formed by the $\gamma$ source and the scintillator (dashed line). }
\label{CCT}
\end{figure}

The \emph{non-proportionality} can be quantified measuring the LY dependency on energy. This esteem can be performed once known both the energy deposited inside the crystal and the corresponding light output. This goal can be accomplished using a method called \emph{Compton Coincidence Technique} (CCT), proposed by \cite{CCT_1}, which exploits the Compton effect to create a monochromatic electron source inside the crystal bulk. This coincidence technique can be used to record events in the scintillator only when a Compton scattered photon is simultaneously detected by a second detector that is located at a fixed angle $\theta$ off from the axis formed by the gamma-ray source and the scintillation detector (see figure \ref{CCT}). 

Fixing both the value of $\theta$ and of the energy of the source gamma ray $E_{\gamma}$ it is possible to fix the energy of the scattered $\gamma$ ray $E'_{\gamma}$ and the Compton electron energy $E_{el}$, as stated by the Compton equation:
\begin{equation} \label{compt}
E_{el}~=~E_{\gamma}~-E'_{\gamma}~=~E_{\gamma}-\frac{E_{\gamma}}{1+\frac{E_{\gamma}}{m_{e}c^2}(1-cos \theta)}
\end{equation}
An additional advantage of this technique is that $E_{el}$ can be easily changed by varying $\theta$ and $E_{\gamma}$. In order to measure precisely $E_{el}$, $E'_{\gamma}$ has to be known with high precision, therefore a high resolution detector was used for this purpose.

The measurements reported in this study have been performed with a CCT setup, using a scintillation crystal coupled to an Hamamatsu R6232-100 PMT and an Ortec HPGe (see figure \ref{CCT}). A $^{22}$Na radioactive source with an activity of $\sim$300~kBq was used, emitting $\gamma$s at 511~keV and 1274~keV. In order to minimize the background counts due to accidental coincidences with $\gamma$s directly going from the source to the HPGe and events from the radioactive decays of $^{40}$K contained in the laboratory walls, the whole setup has been shielded with lead bricks. The signals coming from the scintillator has been preamplified with an Ortec 113 external preamplifier, whose input capacitance has been chosen to have electronics times shorter than the scintillation time of the crystal, thus keeping all the shape characteristic of the scintillation signals. On the other hand, the HPGe signal is first preamplified by the built-in electronics and then transmitted to an amplifier and timing single channel analyser Silena 7216 (AMP-TISCA), providing a digital pulse when the maximum amplitude of the amplified signal falls in a window selected with the potentiometers of this module, correctly calibrated in energy. The signals are acquired with a NI-PXI 5105 12-Bit digitizer, programmed to acquire simultaneously the signals from the scintillator preamplifier and the HPGe amplifier. The acquisition is triggered by the AMP-TISCA digital output, therefore the signals are acquired when the selected energy is detected by the HPGe detector. The number of digitized points for every trigger has been set to ensure the full digitalization of all the coincident signals, usually acquiring a $\sim40$~$\mu$s window with a $\sim60$~MHz sampling frequency. 

\begin{figure}[htbp]
\centering
\includegraphics[width=.6\textwidth,origin=c]{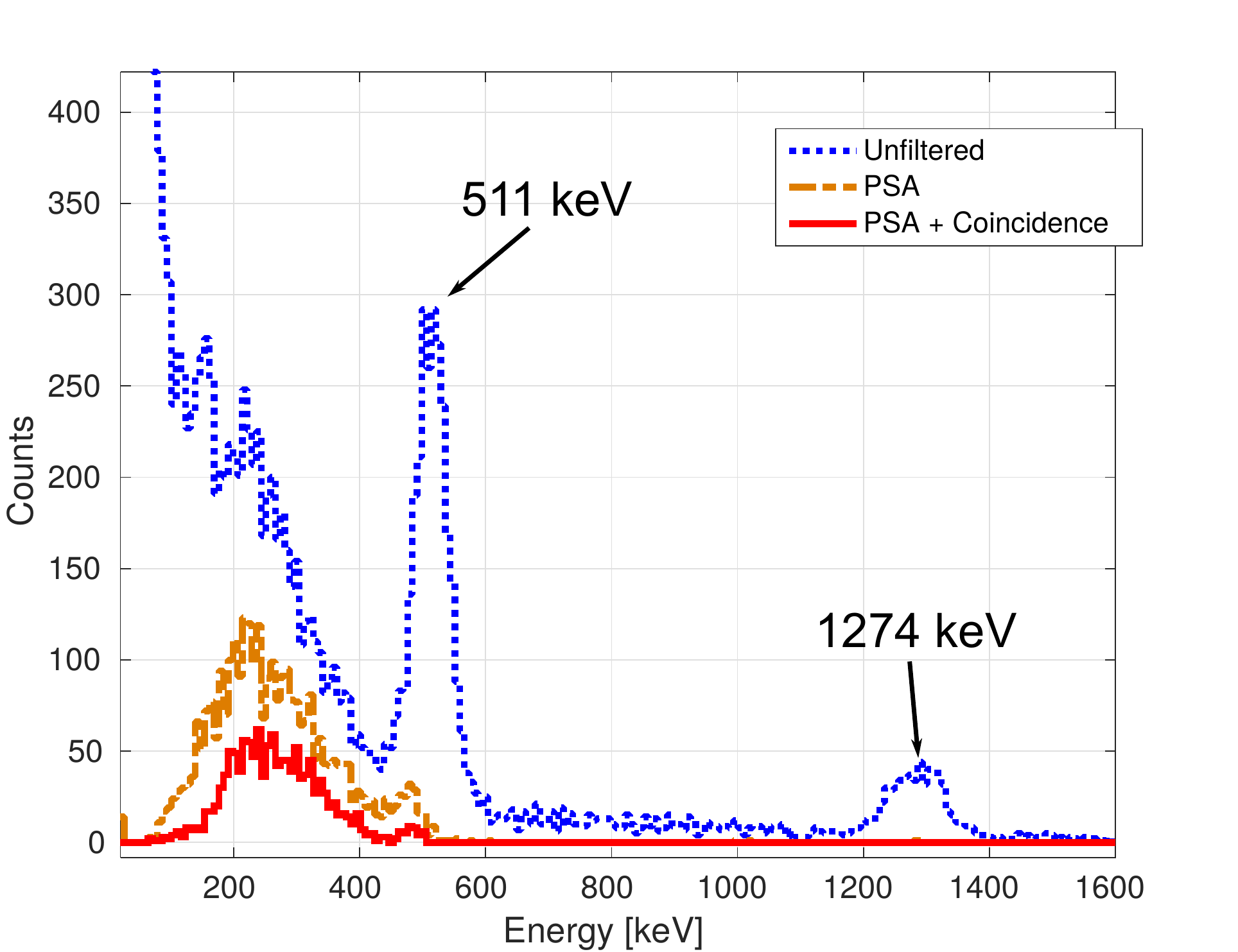}

\caption[Subsequent cuts on CCT scintillator spectrum]{$^{22}$Na spectrum acquired with a CdWO$_4$ scintillator during CCT measurements, looking at the Compton scattering of 1274 keV photons at an angle of 30$^{\circ}$. The combination of coincidence with HPGe signals and the PSA cuts allows to keep only the counts related to the Compton electrons generated inside the crystal bulk. The calibration of the CdWO$_4$ scintillator has been performed with $^{22}$Na, $^{60}$Co and $^{137}$Cs sources.}
\label{SpettroPulitoCCT}
\end{figure}

These measurements have been optimized, in order to exclude noise or bad digitized pulses, applying a pulse shape analysis (PSA) to the scintillator signals. The combination of coincidence technique and PSA cleans up the acquired spectrum, leaving only the searched Compton signals and the spurious coincidences due to background radioactivity. In figure \ref{SpettroPulitoCCT}, it is clearly visible the reduction of the low energy pulses due to spurious coincidences and electronic noise, performed with the combination of coincidence technique and PSA.

\begin{figure}[htbp]
\centering

\includegraphics[width=.45\textwidth,origin=c]{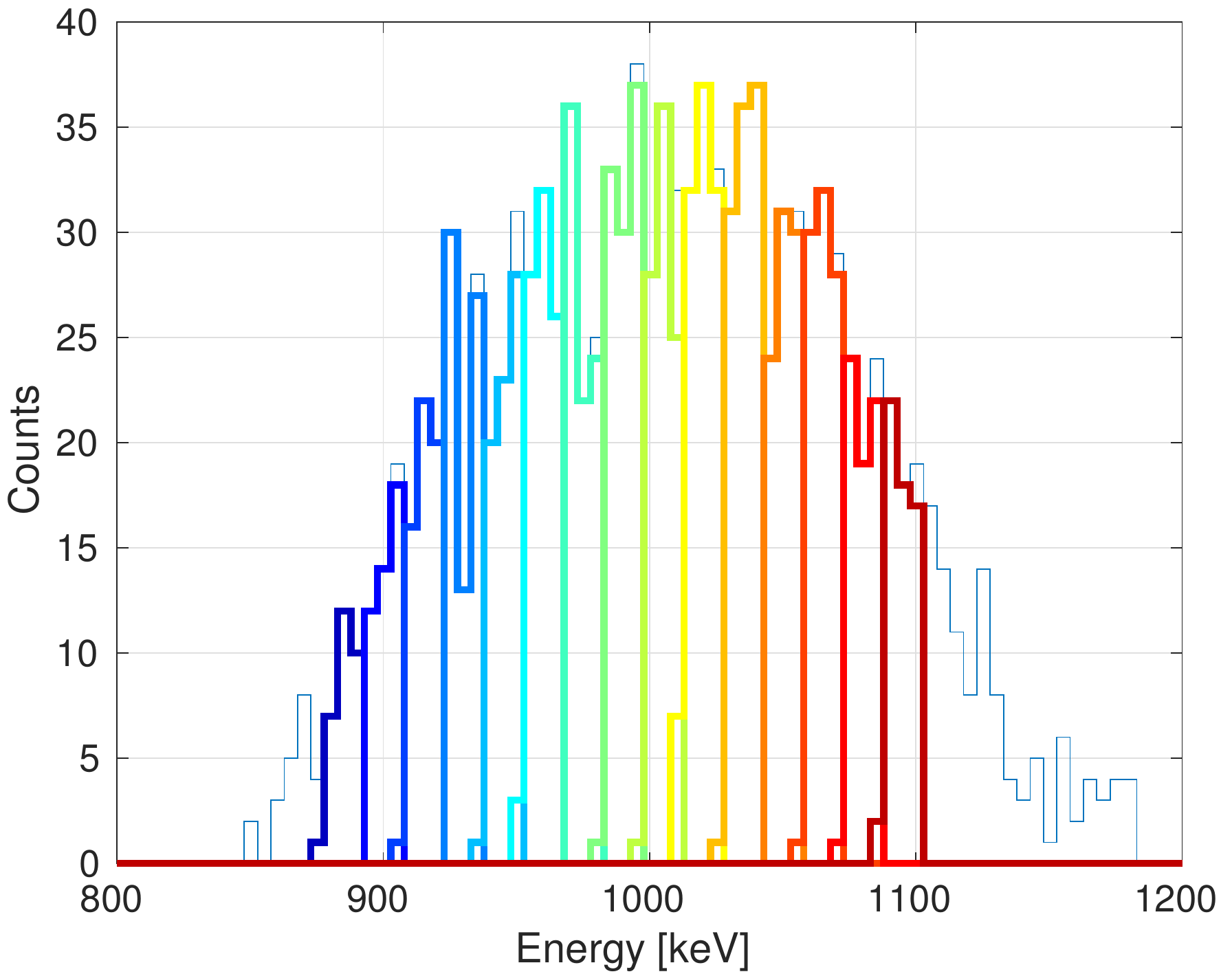}
\qquad
\includegraphics[width=.45\textwidth,origin=c]{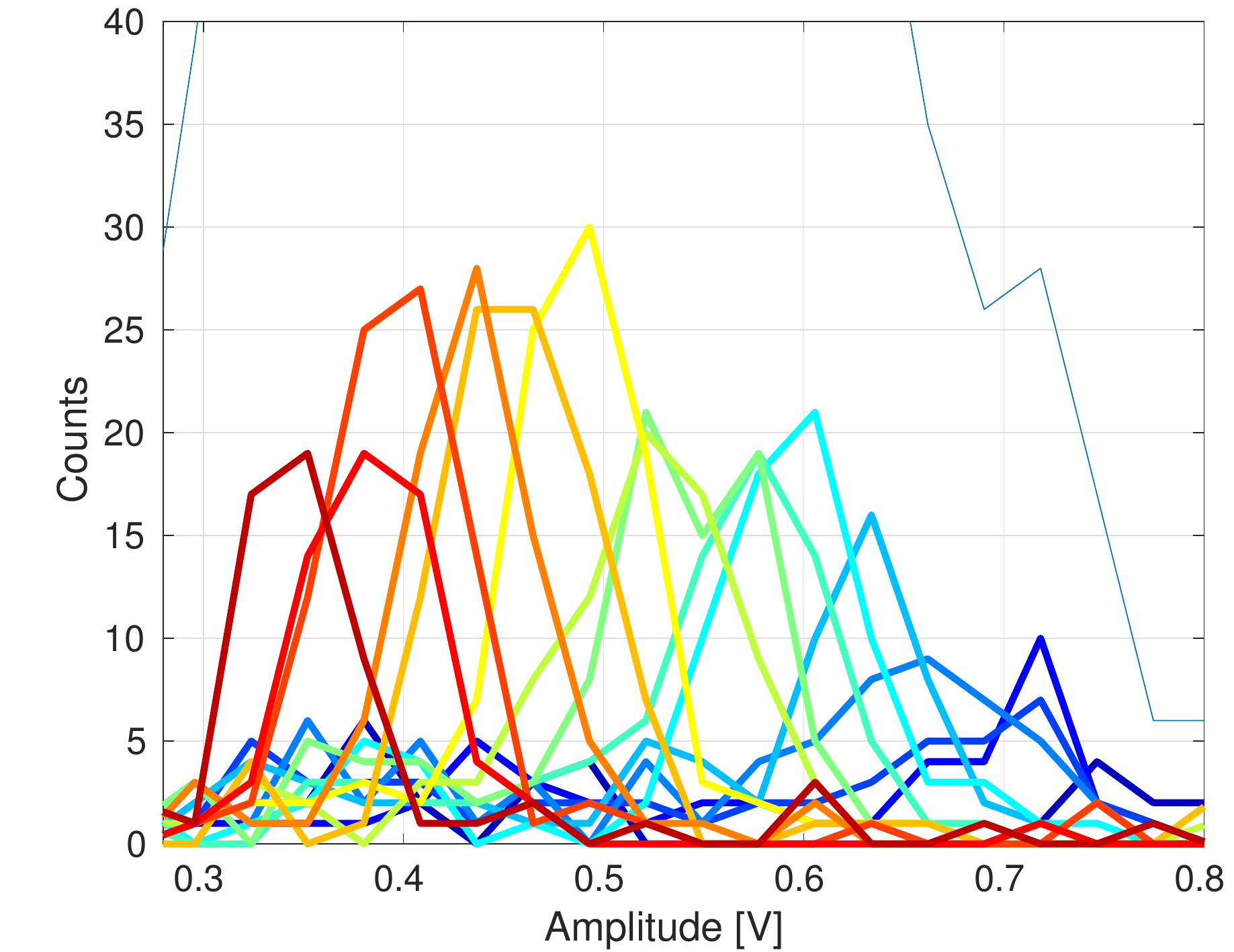}

\caption[Slicing of CCT spectra]{Spectra acquired with the CCT by the HPGe (left), and by the scintillator (right). The event recorded is the Compton scattering of 1274~keV photons at an angle of $\sim$30$^{\circ}$, corresponding to a 900-1100~keV window in the HPGe and to a 170-370~keV window in the scintillator. The different colours correspond to different energies in the HPGe spectrum and the same colour identifies sets of coincident pulses. It is noticeable that higher energy sections in the HPGe correspond to lower energy section in the scintillator, as expected by the Compton scattering events happening inside the crystal. The side bands of the HPGe spectrum are neglected in the analysis, since moving away from the correct scattering energy (~950 keV in the HPGe) the signal becomes lower than the background due to spurious coincidences. As it can be noticed the events on the side region of the window in the scintillator do not present a gaussian shape, because of the increasing background. The scintillator spectrum is not calibrated since the energy is given by $E_{el}~=~E_{\gamma}-E'_{\gamma}$. The resolution measured for the scintillator peaks in the middle of the selected widow are coherent with the values obtained in the calibration measurements, conveniently scaled for the energy obtained by the previous formula.}
\label{SpettriFetteCCT}
\end{figure}

The original design of CCT \cite{CCT_1, CCT_BM} comprehends the use of collimators in order to select more precisely the scattering angle of photons, since both the scattering crystal and the detector subtend a finite solid angle. During these measurements, the physical collimators have been replaced by a software tool, which selects precise energy windows in the HPGe among those signals found in coincidence with the scintillator. As a consequence, multiple measurements were extracted from a single acquisition, optimizing the angular selection and fully exploiting the HPGe resolution (see figure \ref{SpettriFetteCCT}). 

Once selected the scintillator pulses in coincidence with the correct HPGe signals, the light yield can be calculated with the following equation:
\begin{equation}
LY(E_{el})~=~\frac{k \cdot V_{scint}}{E_{\gamma}-E'_{\gamma}}~=~\frac{k \cdot V_{scint}}{E_{el}}
\end{equation}
where $V_{scint}$ is the mean value of the scintillation signal amplitude coincident with the HPGe signal at $E_{\gamma}'$ energy and $k$ is the calibration factor from Volts to photons. In particular, $E_{\gamma}'$ has been evaluated as the weighted mean of the amplitude of the events in the selected energy window in the HPGe spectrum, while $V_{scint}$ as been estimated with an unbinned likelihood fit of the peak in the coincident amplitude spectrum acquired by the scintillator, modelled with a Gaussian subtending a linear background. 

The combination of PSA and coincidences allowed the full characterization of the LY \emph{non-proportionality} of crystals in the range 15-700~keV. Below 15~keV the electronic noise of the scintillator avoids the measurement, while above 700~keV the HPGe background covers the few Compton scattered photons at high angles. The performed measurements are reported in table \ref{CCT_Meas}. 

\begin{table}[htbp]
\centering
\begin{tabular}{|c|c|c|}
\hline
$E_{\gamma}$ [keV] & $\theta$ [deg] & E$_{el}$ [keV]\\
\hline
\multirow{3}*{511} 	& 15		&16\\
					& 20		&29\\
					& 30		&60\\
\hline
\multirow{2}*{1274} 	& 30		&320\\
					& 60		&700\\
\hline
\end{tabular}
\caption[CCT measurements]{CCT measurements performed.}
\label{CCT_Meas}
\end{table}

As Previously said, the LY \emph{non-proportionality} has been measured as a relative variation with respect to the light yield at the 662 keV $\gamma$ line, referred to as \emph{relative light yield} ($R_{LY}$). The obtained variable is given by the following expression:
\begin{equation}
R_{LY}~=~\frac{LY(E_{el})}{LY(662~\text{keV})}~=~\frac{{k} \cdot V_{scint}(E_{el})}{{k} \cdot V_{scint}(662~\text{keV})}\frac{662~keV}{E_{el}}
\end{equation}

\begin{figure}[htbp]
\centering

\includegraphics[width=.5\textwidth,origin=c]{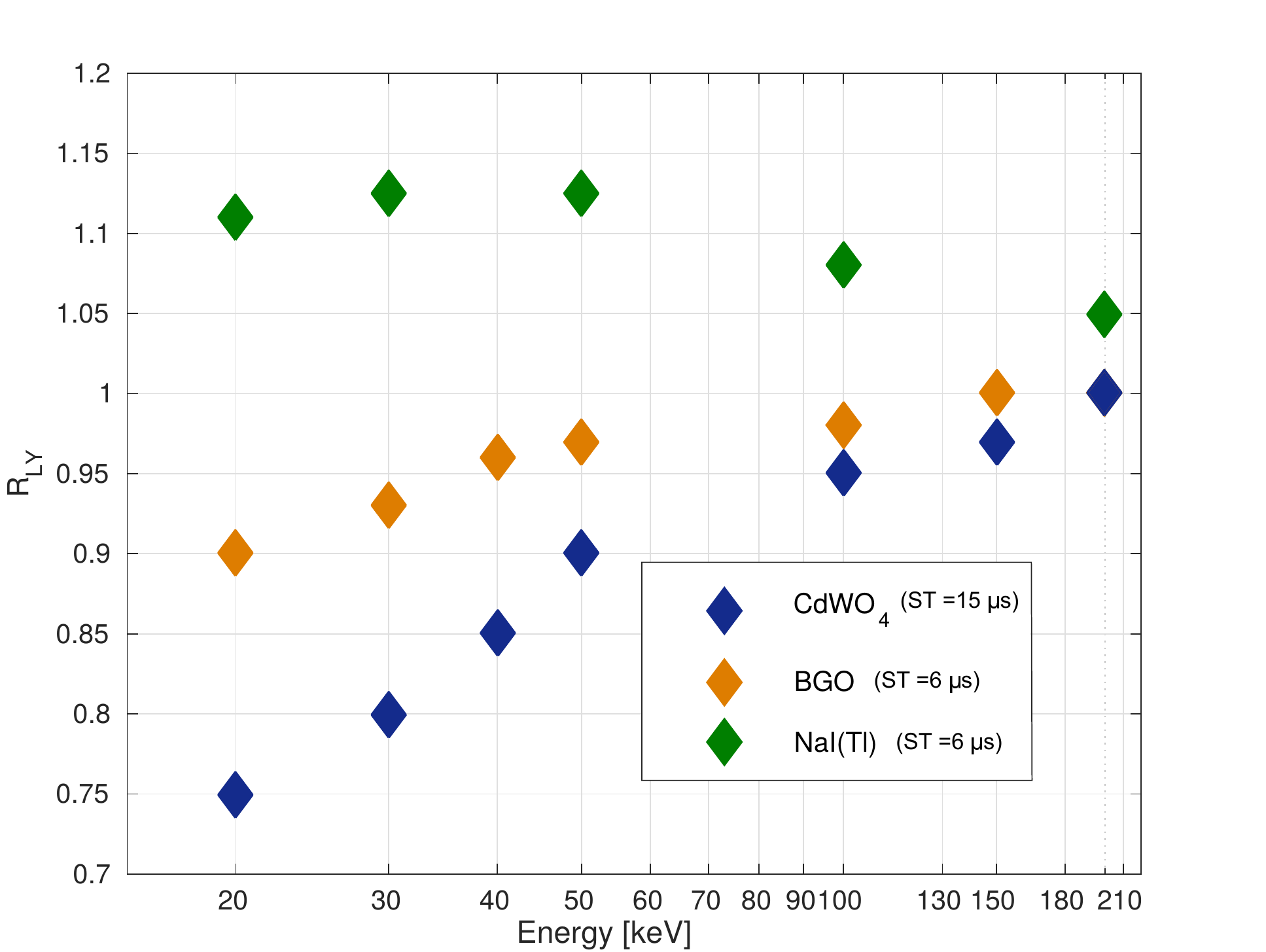}

\caption[Literature non prop]{Literature data on the crystal non proportionality. Data have been taken from from \cite{NonLinRev} and references there in. }
\label{NonPropLit}
\end{figure}

According to the current knowledge \cite{NonLinRev}, both the CdWO$_4$ and BGO crystals should present a negative \emph{non-proportionality}, and their $R_{LY}$ should decrease when the deposited energy lowers. On the other hand, the NaI(Tl) crystal should be positively non-proportional, thus its $R_{LY}$ should increase when the deposited energy lowers (see figure \ref{NonPropLit}). 

\subsection{Shaping time dependency}

\begin{figure}[htbp]
\centering
\includegraphics[width=.5\textwidth,origin=c]{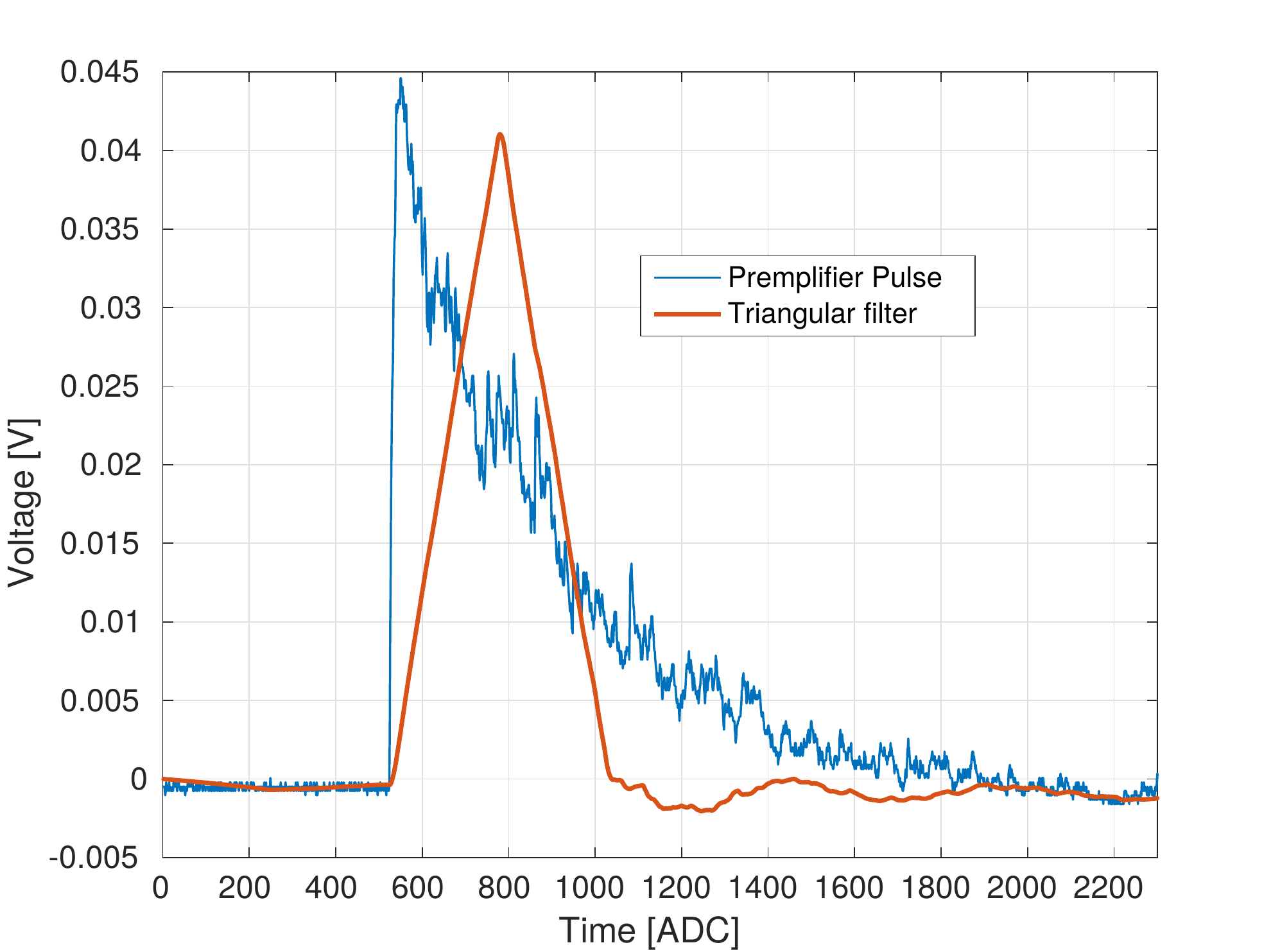}
\caption[Example of triangular shaping]{Example of triangular shaping, superimposed to the preamplifier scintillation pulse of the CdWO$_{4}$ crystal digitized by the ADC at 60~MHz. }
\label{TrapezoidalExample}
\end{figure}

As reported in \cite{NonLinRev} the choice of the shaping time (ST) used to shape the light signals can change the slope of the \emph{non-proportionality} curve. To evaluate the actual effect of ST, it has been chosen to apply a digital pulse processing algorithm (DPP) to the acquired preamplified scintillator pulses. This strategy allows to check the effects of different shaping times, without acquiring multiple measurements for every configuration. The developed algorithm implements the trapezoidal filtering of the acquired pulses and is based on the procedure adopted in \cite{Trapz}. The transfer function of the filter, expressed in the z-transform domain, is:
\begin{equation}
H(z)~=~\frac{1}{R}(1-z^{-R})(1-z^{-(R+M)})\frac{z^{-1}}{(1-z^{-1})^2}(1-\beta z^{-1})
\end{equation}
where $R$ is the number of samples used for rise and decay times, $M$ is the number of samples used for the flat top and $\beta~=~e^{-\frac{T_{sampling}}{T_{preamplifier}}}$ is the factor responsible for the zero-pole cancellation. Since the flat top was affected by noise fluctuations, it has been chosen to use a trapezoidal filter with no flat top, referred to as triangular filter (see figure \ref{TrapezoidalExample}). In addition to the DPP technique, the digitalized preamplified pulses were also integrated numerically. This operation allows to estimate the amplitude of the light pulse without selecting a defined bandwidth, thus including the majority of the scintillation components.

\section{Measured Non-Proportionality}

\begin{figure}[]
\centering

\includegraphics[width=.8\textwidth,origin=c]{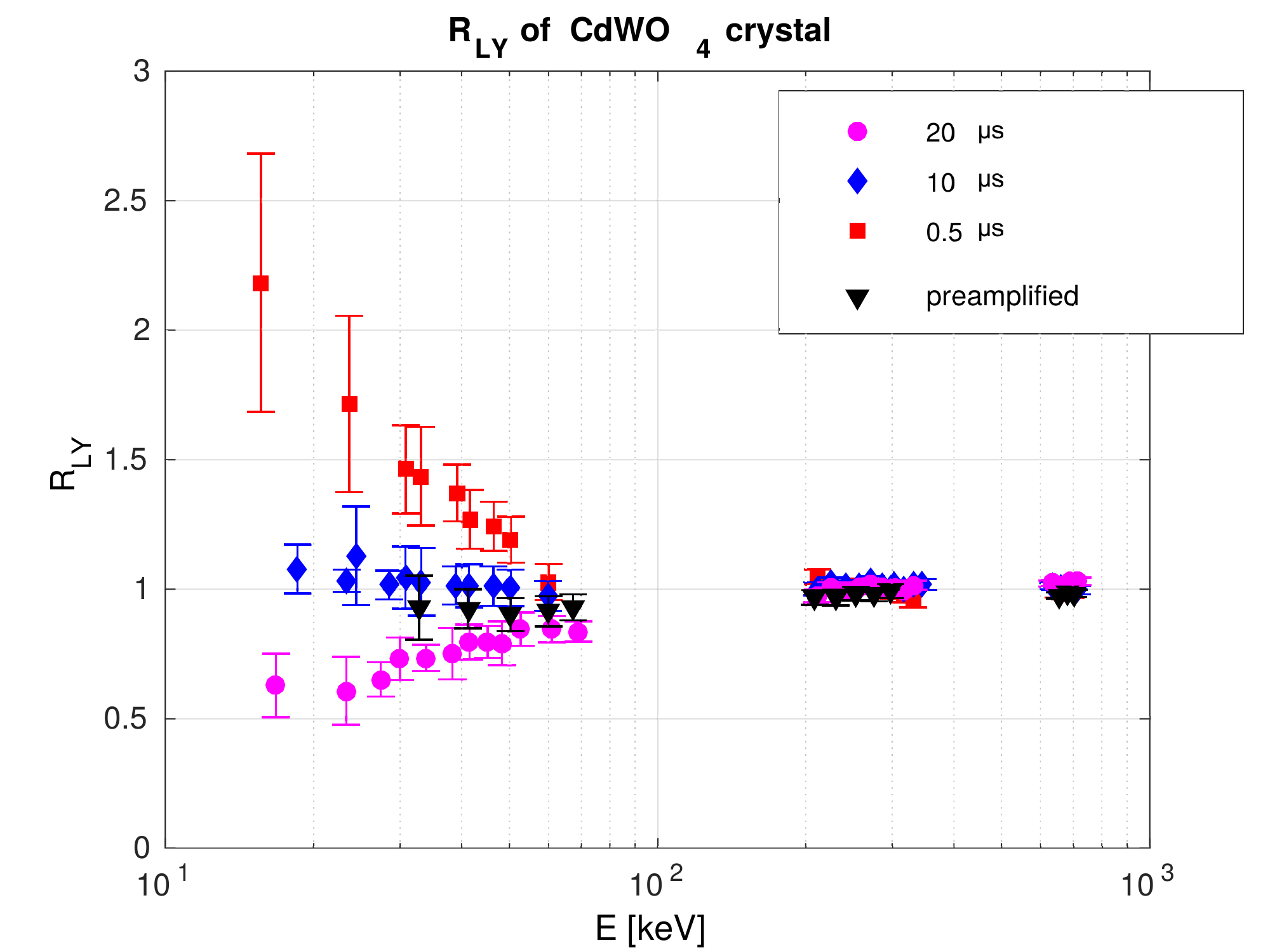}

\caption[measured \emph{non-proportionality} behaviour for a CdWO$_4$ crystal]{Measured \emph{non-proportionality} behaviour for a CdWO$_4$ crystal. The effect of the ST is evident, since it causes the \emph{non-proportionality} to space in a range between 0.5 and 2.5. As expected, integrating the preamplifier signal causes the crystal to loose its \emph{non-proportionality}.}
\label{RLYCdWO}
\end{figure}

\begin{figure}[htbp]
\centering
\includegraphics[width=.7\textwidth,origin=c]{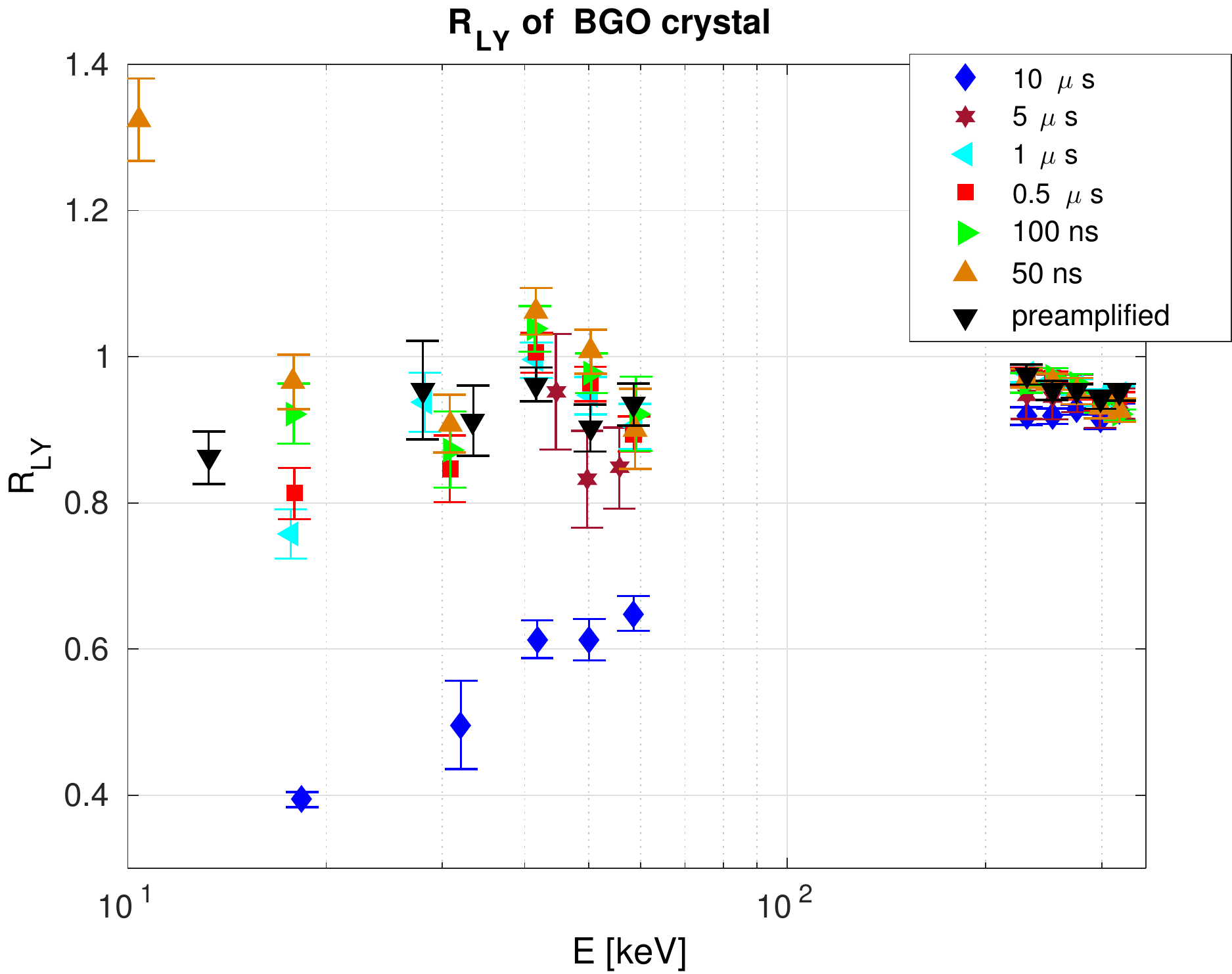}
\caption[measured \emph{non-proportionality} for BGO crystal]{Measured \emph{non-proportionality} behaviour for a BGO crystal. Since the scintillation process of this crystal is quicker than the CdWO$_{4}$, all the ST$>$50~ns give the same effects, and a longer ST (10~$\mu s$) return a negative R$_{LY}$. At high energies the R$_{LY}$ is lower than 1, because the small size of the used BGO crystal (2x2x2cm$^3$) did not allow the complete absorption of the electrons energy. }
\label{RLYBGO}
\end{figure}

\begin{figure}[htbp]
\centering
\includegraphics[width=.7\textwidth,origin=c]{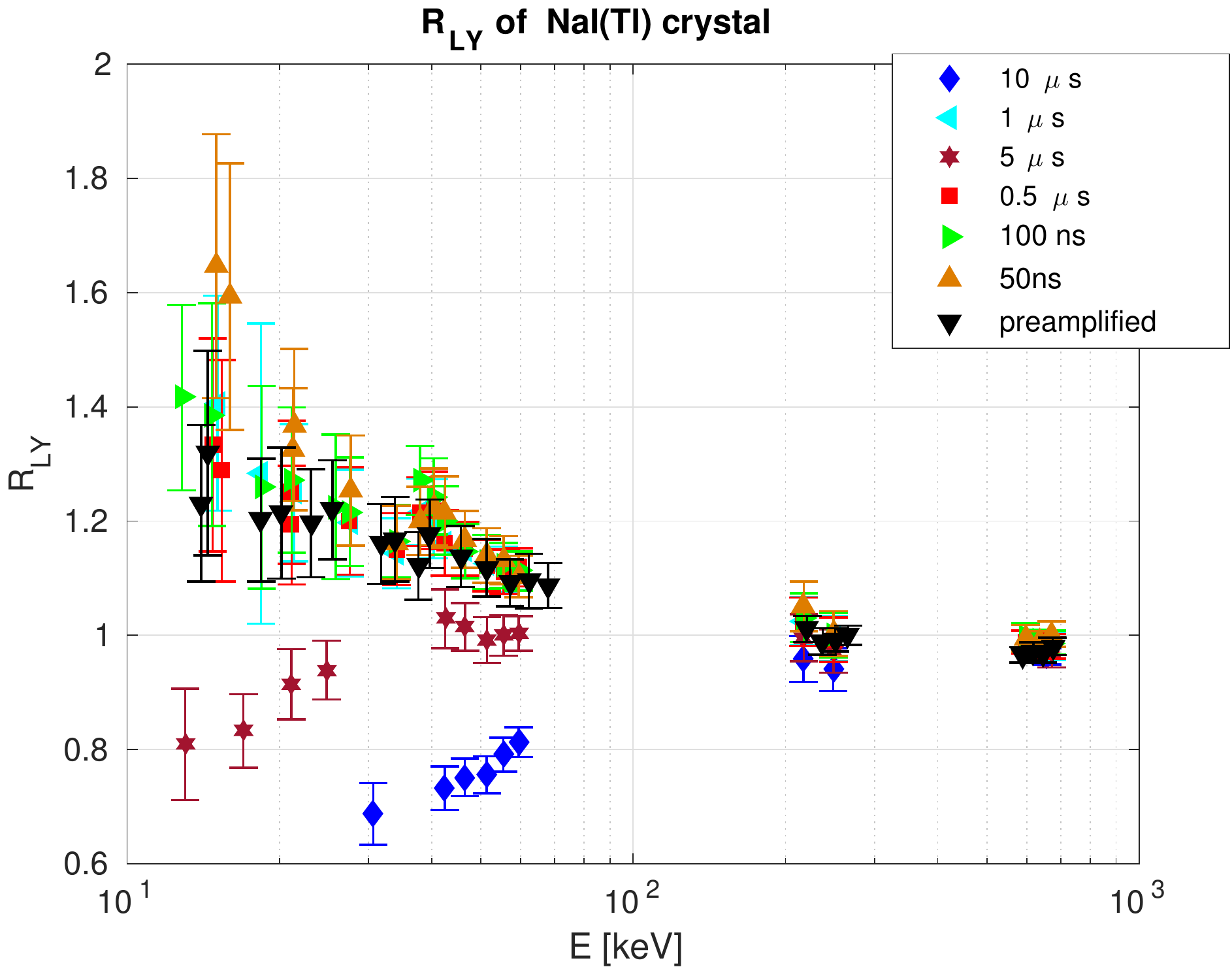}
\caption[measured \emph{non-proportionality} for NaI crystal]{Measured \emph{non-proportionality} for NaI(Tl) crystal. ST of 10$\mu s$ and 5$\mu s$ cause the observed behaviour to change from the literature studies (figure \ref{NonPropLit}).}
\label{RLYNaI}
\end{figure}

The results obtained for the \emph{non-proportionality} behaviour with different STs, for CdWO$_4$, BGO and NaI(Tl) crystals are reported respectively in figure \ref{RLYCdWO}, \ref{RLYBGO} and \ref{RLYNaI}. From these data it can be inferred that the \emph{non-proportionality} of scintillation crystals is determined by the ST used to process the light signal. Choosing a longer shaping time, in fact, causes all the studied crystals to be negatively non-proportional, while choosing a shorter shaping time causes the same crystals to be positively non-proportional. The transition between negative and positive deviation is different for each studied crystal, and it is related to the time characteristics of its scintillation. For example, on one hand the CdWO$_4$, which has a main scintillation time constant of $\sim14$~$\mu$s, shows negative \emph{non-proportionality} for ST$>$10~$\mu s$. On the other hand the BGO, which has a main time constant of $\sim300$~ns, shows negative deviations for ST$>$50~ns. 

The obtained results show a clear correlation between the \emph{non-proportionality} behaviour with energy and the ST chosen to filter the light pulses. The observed behaviour can be explained by a dependence with energy of the involved scintillation component: faster components are predominant at lower energies, while higher energy events are characterized by longer scintillation times. This can be ascribed to an eventual enrolment mechanism of the activation sites in the crystal lattice: on one hand, the deposition of small energies activates just few sites characterized by a short decay time; on the other hand, an event with higher energy causes the activation of different recombination sites, with both short and long characteristic times. As a consequence it could be inferred a sort of hierarchy between the recombination sites, ruled by the energy of the activation event and by its ionizing potential.  

\section{Summary}
A detailed study of the \emph{non-proportionality} behaviour with energy of three different scintillation crystals (CdWO$_{4}$, BGO and NaI) has been performed. The relation between \emph{non-proportionality} and ST has been characterized by the means of a DPP algorithm, which allowed to test different integration times on the same measurement. The obtained results suggest that the \emph{non-proportionality} in light yield of scintillation crystals is not an intrinsic characteristics of the crystal itself: as a matter of fact, it depends on which part of the scintillation spectrum is considered in the pulse processing, since different scintillation components are selected by different STs. 

In the framework of high resolving scintillation detectors, this evidence can be used to tune the analysis strategy for the light pulses, optimizing the spectrometric performances of these crystals. On the other hand, this peculiar characteristics has to be further investigated, in order to enlighten the hidden characteristics of the scintillation mechanism of crystals.


\begin{thebibliography}{9}

\bibitem{NonLinRev}
M. Moszyński and A. Syntfeld-Każuch and L. Swiderski and M. Grodzicka and J. Iwanowska and P. Sibczyński and T. Szczęśniak, \emph{Energy resolution of scintillation detectors}, \emph{Nucl. Instrumention and Methods in Physics Research A} {\bf 805} (2016) 25-35.

\bibitem{NonPropEResolutionLimits}
P. Dorenbos and J. T. M. de Haas and C. W. E. van Eijk, \emph{Non-proportionality in the scintillation response and the energy resolution obtainable with scintillation crystals}, \emph{IEEE Transactions on Nuclear Science} {\bf 42} (1995) 2190-2202.

\bibitem{NonPropEResolution}
J. D. Valentine and B. D. Rooney and J. Li, \emph{The light yield nonproportionality component of scintillator energy resolution}, \emph{IEEE Transactions on Nuclear Science} {\bf 45} (1998) 512-517.

\bibitem{CCT_1}
John D. Valentine, Brian D. Rooney, \emph{Design of a Compton spectrometer experiment for studying scintillator non-linearity and intrinsic energy resolution}, \emph{Nuclear Instruments and Methods in Physics Research A} {\bf 353} (1994) 37-40.

\bibitem{CCT_BM}
John D. Valentine and Brian D. Rooney, \emph{Benchmarking the Compton Coincidence Technique for Measuring Electron Response Non-Proportionality in Inorganic Scintillators}, \emph{IEEE Transactions on Nuclear Science} {\bf 43} (1996) 1271-1276.

\bibitem{GammaFromEl}
Brian D. Rooney and John D. Valentine, \emph{Scintillator Light Yield Nonproportionality : Calculating Photon Response Using Measured Electron Response}, \emph{IEEE Transactions on Nuclear Science} {\bf 44} (1997) 509-516.

\bibitem{Knoll}
Glenn F. Knoll, \emph{Radiation Detection and Measurement},
John Wiley \& Sons, Inc. (2010).

\bibitem{BGOData}
N. Tsuchida, M. Ikeda, T. Kamae, M. Kokubun, \emph{Temperature dependence of gamma-ray excited scintillation time profile and light yield of GSO, YSO, YAP and BGO}, \emph{Nuclear Instruments and Methods in Physics Research Section A} {\bf 385} (1997) 290-298.

\bibitem{NaIData}
C. Cuesta, M.A. Oliván, J. Amaré, S. Cebrián, E. García, C. Ginestra, M. Martínez, Y. Ortigoza, A. Ortiz de Solórzano, C. Pobes, J. Puimedón, M.L. Sarsa, J.A. Villar, P. Villar, \emph{Slow scintillation time constants in NaI(Tl) for different interacting particles}, \emph{Optical Materials} {\bf 36} (2013) 316-320.

\bibitem{DifferentST}
M. Moszyński, W. Czarnacki, W. Klamra, M. Szawlowski, P. Schotanus, M. Kapusta, \emph{Intrinsic energy resolution of pure NaI studied with large area avalanche photodiodes at liquid nitrogen temperatures}, \emph{Nucl. Instruments and Methods in Physics Research Section} {\bf 505} (2003) 63.

\bibitem{MurrMayer}
A. Meyer and R. B. Murray, \emph{Scintillation Response of Activated Ionic Crystals to Charged Particles}, \emph{IRE Transactions on Nuclear Science} {\bf 7} (1960) 22-25.


\bibitem{Trapz}
Z. Guzik and T. Krakowski, \emph{Algorithms for digital $\gamma$-ray spectroscopy}, \emph{Nukleonika} {\bf 58} (2013) 333-338.


\end{thebibliography}
\end{document}